\documentclass[
reprint,
superscriptaddress,
amsmath,
amssymb,
aps
]{revtex4-2}

\usepackage{orcidlink}
\usepackage{amsmath}
\usepackage{hyperref}
\usepackage{mathtools}
\usepackage{color}
\usepackage{soul}
\sethlcolor{yellow}

\def\Lie{{\pounds}}

\begin{document}

\title{Reflecting boundary conditions in numerical relativity as a model for black hole echoes}

\author{Conner Dailey \orcidlink{0000-0003-2488-3461}
}
\email[Corresponding Author: ]{cdailey@pitp.ca}
\affiliation{Waterloo Centre for Astrophysics, University of Waterloo, Waterloo, ON, N2L 3G1, Canada}
\affiliation{Department of Physics and Astronomy, University of Waterloo, Waterloo, Canada}
\affiliation{Perimeter Institute for Theoretical Physics, Waterloo, Canada}

\author{Niayesh Afshordi \orcidlink{0000-0002-9940-7040}}
\affiliation{Waterloo Centre for Astrophysics, University of Waterloo, Waterloo, ON, N2L 3G1, Canada}
\affiliation{Department of Physics and Astronomy, University of Waterloo, Waterloo, Canada}
\affiliation{Perimeter Institute for Theoretical Physics, Waterloo, Canada}

\author{Erik Schnetter \orcidlink{0000-0002-4518-9017}}
\affiliation{Perimeter Institute for Theoretical Physics, Waterloo, Canada}
\affiliation{Department of Physics and Astronomy, University of Waterloo, Waterloo, Canada}
\affiliation{Center for Computation \& Technology, Louisiana State University, Baton Rouge, Louisiana, USA}

\date{2023-01-12}

\begin{abstract}
Recently, there has been much interest in black hole echoes, based on the idea that there may be some mechanism (e.g., from quantum gravity) that waves/fields falling into a black hole could partially reflect off of an interface before reaching the horizon. There does not seem to be a good understanding of how to properly model a reflecting surface in numerical relativity, as the vast majority of the literature avoids the implementation of artificial boundaries, or applies transmitting boundary conditions. Here, we present a framework for reflecting a scalar field in a fully dynamical spherically symmetric spacetime, and implement it numerically. We study the evolution of a wave packet in this situation and its numerical convergence, including when the location of a reflecting boundary is very close to the horizon of a black hole. This opens the door to model exotic near-horizon physics within full numerical relativity. 
\end{abstract}

\maketitle

\section{Introduction}

The recent discovery of gravitational waves (GWs) has led to the new field of GW astronomy, opening novel windows into some of the deepest physical mysteries in our universe. Chief amongst these is the true nature of black hole horizons, which appear to have both temperature and entropy, if one applies laws of quantum mechanics to the fields in their vicinity\cite{Hawking1975,Bekenstein1972}. Yet, according to Einstein's theory of General Relativity (GR), horizons comprise nothing but empty space. Even more problematic is when the thermal nature of black holes (i.e. Hawking radiation) leads to their evaporation, leaving us to wonder what happens to the information behind the horizon that cannot escape due to relativistic causality. This is known as the black hole information paradox\cite{Hawking1976,Mathur_2009}. Given that GW astronomy can now probe into the vicinity of black hole horizons, it is natural to ask whether it can give us an empirical window into resolving these theoretical mysteries. In particular, one may wonder whether quantum effects near the horizon would lead to echoes in observations that would have been absent in GR with classical horizons\cite{Abedi:2016hgu}.

Motivated by this,  there has been a recent interest in  modelling black hole echoes phenomenologically\cite{Oshita:2019sat,Wang:2018gin,Ikeda:2021uvc,Wang:2019rcf,Abedi:2018npz,Abedi:2020sgg,Abedi:2016hgu,Burgess:2018pmm,Cardoso:2019apo}.
The idea behind black hole echoes is that there may be some mechanism, resulting from quantum phenomena for example, by which wave packets falling into a black hole could partially reflect off of an interface before reaching the horizon, thus resulting in a potentially detectable effect. In the context of GW astronomy, this has opened a novel window to study exotic near-horizon physics phenomenologically, from current and future observations \cite{Abedi:2018npz,Abedi:2020sgg,Abedi:2016hgu,Cardoso:2019apo}. However, interpreting this data requires modelling strong gravitational systems in numerical relativity, where there does not currently seem to be a good understanding of how to properly implement a reflecting surface (but see \cite{Ma:2022xmp} for an approximate method and \cite{Danielsson:2021ykm} for an adjacent approach to boundaries). There has also been some recent skepticism that black hole echoes can be detected at all due to the formation of an apparent horizon before the reflection occurs \cite{GuoMathur}, and having a proper way to model reflecting surfaces allows for rigorous testing beyond order-of-magnitude estimates.


One way to implement a reflecting surface is to have an artificial boundary in the domain of the numerical problem where the surface is to be located, and impose a set of boundary conditions (BCs) there. This requires an understanding of the full initial boundary value problem (IBVP) for a particular framework of numerical relativity, which the majority of  studies tend to avoid by imposing no boundaries or by having boundaries out of causal contact with the initial conditions. This work contains some examples of the full spherically symmetric IBVP in numerical relativity with a scalar field. Of course there are no GWs in spherical symmetry, but see Appendix~\ref{App:GWs} for how this can be done analagously in cylindrical symmetry.

The generic way of treating the BCs in hyperbolic systems of partial differential equations is to identify the characteristic modes and speeds of the system, and for each mode that has a characteristic speed leading it out of the domain, no BCs are applied, while all of the modes propagating into the domain require a boundary treatment. It is then advantageous to pick a formulation of Einstein's equations where the characteristic modes and speeds are simple and well-behaved. Here, the Einstein-Christoffel (EC) formulation is used due to its simple characteristic structure, symmetric hyperbolicity, and previously successful numerical implementation \cite{EC_Formulation}. 

In Section~\ref{Sec:EC_System}, we describe the EC system and the scalar field equations adapted from \cite{EC_Formulation}. In Section~\ref{Sec:BCs}, we derive BCs based on conservation laws while Section~\ref{Sec:ICs} dictates the initial conditions we consider. Section~\ref{Sec:Numerics} describes our numerical implementation using summation by parts (SBP) operators and BCs implemented with simultaneous approximation terms (SATs). Finally, we present our results for several reflecting IBVPs in Section~\ref{Sec:Results}, discuss near-horizon boundaries, and discuss the evolution of the apparent horizon as a response to \cite{GuoMathur}.

\section{Theoretical Framework for the Einstein-Christoffel system}\label{Sec:EC_System}

In general, the physical metric $g_{\mu\nu}$ in spherical symmetry can be expressed with respect to the spherical coordinates $x^\alpha=(t,r,\theta,\varphi)$, with the line element:
\begin{align}
    ds^2 = & -\alpha^2dt^2+\gamma_{rr}(dr+\beta^r dt)^2+\gamma_{\theta\theta}d\Omega^2\,,
\end{align}
where $\alpha$ is the lapse, $\beta^r$ is the shift, and $\gamma_{ij}$ are elements of the 3-metric. The extrinsic curvature is defined as
\begin{equation}
    K_{ij} = -\frac{1}{2\alpha}\left(\partial_t-\Lie_\beta\right)\gamma_{ij}\,,
\end{equation}
where $\Lie_\beta$ represents the Lie derivative with respect to the shift vector. In order to make the evolution equations first order in space, the functions
\begin{equation}
    f_{kij} \equiv \Gamma_{(ij)k}+\gamma_{ki}\gamma^{lm}\Gamma_{[lj]m}+\gamma_{kj}\gamma^{lm}\Gamma_{[li]m}\,,\label{Eq:fijk}
\end{equation}
are defined. Here, $\Gamma^k_{\ ij}$ are the Christoffel symbols associated with the 3-metric. The EC framework leaves the densitized lapse $\tilde\alpha\equiv\alpha/(\sqrt{\gamma_{rr}}\gamma_{\theta\theta})$ and the shift as freely specifiable gauge functions. It is understood that wherever $\alpha$ may appear in the evolution equations, it is set with $\alpha = \tilde\alpha\sqrt{\gamma_{rr}}\gamma_{\theta\theta}$. While we present a general formalism in what follows, we shall come back to our particular choice of gauge in Section (\ref{sec:gauge}). 


\subsection{The Einstein-Christoffel vacuum equations}

The evolution equations in the EC system in spherical symmetry were introduced in detail in \cite{EC_Formulation}, which we modify to the following form:
\begin{widetext}
\begin{align}
    \partial_t\gamma_{rr} -\beta^r\partial_r\gamma_{rr}  = &\  2\gamma_{rr}\partial_r\beta^r - 2\alpha K_{rr}\,, \label{Eq.EC_r}\\
    \partial_t\gamma_{\theta\theta} -\beta^r\partial_r\gamma_{\theta\theta} = & -2\alpha K_{\theta\theta}\,, \label{Eq.EC_theta}\\
    \partial_t K_{rr}- \beta^r\partial_rK_{rr} +\alpha\mathcal{D}^rf_{rrr} = &\  3\frac{\alpha f_{rrr}^2}{\gamma_{rr}^2} -6\frac{\alpha f_{r\theta\theta}^2}{\gamma_{\theta\theta}^2} - \frac{\alpha K_{rr}^2}{\gamma_{rr}} + 2\frac{\alpha K_{rr}K_{\theta\theta}}{\gamma_{\theta\theta}} - 10\frac{\alpha f_{rrr}f_{r\theta\theta}}{\gamma_{rr}\gamma_{\theta\theta}}- \frac{\alpha f_{rrr}\partial_r\ln\tilde\alpha}{\gamma_{rr}}\nonumber\\&  - \alpha\, (\partial_r\ln\tilde\alpha)^2 -\alpha\,\partial_r^2\ln\tilde\alpha + 2 K_{rr}\partial_r\beta^r\,,\label{Eq.EC_Krr}\\
    \partial_t K_{\theta\theta} - \beta^r\partial_rK_{\theta\theta} +\alpha\mathcal{D}^rf_{r\theta\theta} = &\ \alpha + \frac{\alpha K_{rr}K_{\theta\theta}}{\gamma_{rr}}+\frac{\alpha f_{rrr}f_{r\theta\theta}}{\gamma_{rr}^2} -4\frac{\alpha f_{r\theta\theta}^{2}}{\gamma_{rr}\gamma_{\theta\theta}} -\frac{\alpha f_{r\theta\theta}\partial_r\ln\tilde\alpha}{\gamma_{rr}}\,,\label{Eq.EC_Ktt}\\
    \partial_t f_{rrr} - \beta^r\partial_rf_{rrr} +\alpha\partial_rK_{rr} = &\ 12\frac{\alpha \gamma_{rr}K_{\theta\theta}f_{r\theta\theta}}{\gamma_{\theta\theta}^2} -\frac{\alpha K_{rr}f_{rrr}}{\gamma_{rr}} - 10\frac{\alpha K_{rr}f_{r\theta\theta}}{\gamma_{\theta\theta}} - 4\frac{\alpha K_{\theta\theta}f_{rrr}}{\gamma_{\theta\theta}}- \alpha K_{rr}\partial_r\ln\tilde\alpha\nonumber\\ &  - 4\frac{\alpha\gamma_{rr} K_{\theta\theta}\partial_r\ln\tilde\alpha}{\gamma_{\theta\theta}}+ 3f_{rrr}\partial_r\beta^r + \gamma_{rr}\partial_r^2\beta^r\,,\label{Eq.EC_frrr}\\ 
    \partial_t f_{r\theta\theta} - \beta^r\partial_rf_{r\theta\theta}+\alpha\partial_rK_{\theta\theta} = &\ 2\frac{\alpha K_{\theta\theta} f_{r\theta\theta}}{\gamma_{\theta\theta}}-\frac{\alpha K_{\theta\theta} f_{rrr}}{\gamma_{rr}} - \alpha K_{\theta\theta}\partial_r\ln\tilde\alpha 
    + f_{r\theta\theta}\partial_r\beta^r\,.\label{Eq.EC_frtt}
\end{align}
\end{widetext}
There are a few differences in this form of the equations compared to those presented in \cite{EC_Formulation}. First, we write the equations explicitly in terms of the operator $\mathcal{D}^i$, which acts on a 3-covector $f_i$ as ${\mathcal{D}^if_i=(\sqrt{\gamma})^{-1}\partial^i(\sqrt{\gamma}f_i)}$ (reasons for this will be given in Section~\ref{Sec:SBP}). 
Also, instead of defining transverse elements like $\gamma_T=\gamma_{\theta\theta}/r^2$, we keep the evolution equations in a form that does not explicitly depend on the coordinates. The principal part of this system makes up the left hand side and source-like terms make up the right hand side, which suggests the simple characteristic structure of this formulation of Einstein's vacuum equations. We should note that there is only gauge dynamics possible in the above system until we add stress-energy terms, which we do next.

\subsection{Scalar Field and Source Terms}

Scalar fields of mass $m$ are subject to the Klein-Gordon equation $\nabla^\mu\nabla_\mu\phi-m^2\phi=0$. This equation can be reduced to a first order system and keep the same characteristic structure as the EC system with the definition of two auxiliary variables: $\psi_r\equiv\partial_r\phi$ and ${\Pi\equiv-(\partial_t\phi-\beta^r\partial_r\phi)/\alpha}$. The evolution equations then become 
\begin{widetext}
\begin{align}
\partial_t \phi-\beta^r\partial_r\phi&=   -\alpha\Pi\,,\label{Eq-KG1}\\
    \partial_t \psi_r -\beta^r\partial_r\psi_r+\alpha\partial_r\Pi  &=\alpha\left(-\frac{f_{rrr}}{\gamma_{rr}}+2\frac{f_{r\theta\theta}}{\gamma_{\theta\theta}}-\partial_r\ln\tilde\alpha\right)\Pi + \partial_r\beta^r\psi_r\,,\label{Eq-KG2}\\
    \partial_t \Pi -\beta^r\partial_r\Pi+\alpha\mathcal{D}^r\psi_r &= \alpha\left(\frac{K_{rr}}{\gamma_{rr}}+2\frac{K_{\theta\theta}}{\gamma_{\theta\theta}}\right)\Pi + \frac{\alpha}{\gamma_{rr}}\left(\frac{f_{rrr}}{\gamma_{rr}}-6\frac{f_{r\theta\theta}}{\gamma_{\theta\theta}}-\partial_r\ln\tilde\alpha\right)\psi_r+m^2\alpha\phi\,.\label{Eq-KG}
\end{align}
\end{widetext}
With the unit normal to the spatial foliation $n_\mu$, the stress-energy tensor $T^{\mu\nu}$ is decomposed into the energy density $\rho = n_\mu n_\nu T^{\mu\nu}$, the momentum density ${S_i = - \gamma_{i\mu} n_\nu T^{\mu\nu}}$, the trace $T=T^\mu_{\ \mu}$, and the spatial stress $S_{ij}=\gamma_{i\mu}\gamma_{j\nu}T^{\mu\nu}$. In spherical symmetry the nonzero elements for the Klein-Gordon field are explicitly
\begin{align}
    \rho&= \left(\Pi^2+\psi_r\psi^r+m^2\phi^2\right)/2\,,\\
    S_r  &= \Pi\psi_r\,,\\
    T &= \Pi^2-\psi_r\psi^r-2m^2\phi^2\,,\\
    S_{rr} &= \gamma_{rr}\left(\Pi^2+\psi_r\psi^r-m^2\phi^2\right)/2\,,\\
    S_{\theta\theta} &= \gamma_{\theta\theta}\left(\Pi^2-\psi_r\psi^r-m^2\phi^2\right)/2\,.
\end{align}
These elements modify the right hand side Eqns.~(\ref{Eq.EC_Krr}-\ref{Eq.EC_frrr}) in this framework as
\begin{align}
    \partial_t K_{rr}&= \cdots +4\pi\alpha(\gamma_{rr}T-2S_{rr})\,,\\
    \partial_t K_{\theta\theta}  &=\cdots +4\pi\alpha(\gamma_{\theta\theta}T-2S_{\theta\theta})\,,\\
    \partial_t f_{rrr}  &=\cdots + 16\pi\alpha\gamma_{rr}S_r\,,
\end{align}
and this system is subject to a set of five constraints:
\begin{align}
    C\equiv\frac{\partial_rf_{r\theta\theta}}{\gamma_{rr}\gamma_{\theta\theta}}+\frac{7}{2}\frac{f_{r\theta\theta}^2}{\gamma_{rr}\gamma_{\theta\theta}^2}-\frac{f_{rrr}f_{r\theta\theta}}{\gamma_{rr}^2\gamma_{\theta\theta}}-\frac{1}{2}\frac{K_{\theta\theta}^2}{\gamma_{\theta\theta}^2}\nonumber\\-\frac{1}{2\gamma_{\theta\theta}}-\frac{K_{rr}K_{\theta\theta}}{\gamma_{rr}\gamma_{\theta\theta}}+4\pi\rho&=0\label{Eq.C}\,,\\
    C_r\equiv\frac{\partial_rK_{\theta\theta}}{\gamma_{\theta\theta}}-\frac{K_{\theta\theta}f_{r\theta\theta}}{\gamma_{\theta\theta}^2}-\frac{K_{rr}f_{r\theta\theta}}{\gamma_{rr}\gamma_{\theta\theta}}+4\pi S_r&=0\label{Eq.Cr}\,,\\
    C_{rrr}\equiv\partial_r\gamma_{rr}+8\frac{\gamma_{rr}f_{r\theta\theta}}{\gamma_{\theta\theta}}-2f_{rrr}&=0\,,\label{Eq.fr}\\[0.1cm]
    C_{r\theta\theta}\equiv\partial_r\gamma_{\theta\theta}-2 f_{r\theta\theta}&=0 \,,\label{Eq.f_theta}\\[0.2cm]
    C_\phi\equiv\partial_r\phi-\psi_r&=0 \,.\label{Eq:C_phi}
\end{align}
Eqns.~(\ref{Eq.C}) and (\ref{Eq.Cr}) are the Hamiltonian and Momentum constraints respectively, Eqns.~(\ref{Eq.fr}) and (\ref{Eq.f_theta}) come from definition Eq.~(\ref{Eq:fijk}), and Eq.~(\ref{Eq:C_phi}) was defined at the beginning of this section.


The characteristic modes for the complete system of equations include a set of three modes that propagate along the timelike normal to the foliation with characteristic speed $-\beta^r$~:
\begin{align}
    U^0_r=\gamma_{rr}\,,\quad U^0_\theta=\gamma_{\theta\theta}\,,\quad U^0_\phi=\phi\,,
\end{align}
and a set of six modes that propagate along the light cone with characteristic speeds $c_\pm \equiv -\beta^r\pm\alpha/\sqrt{\gamma_{rr}}$~:
\begin{align}
    U^\pm_r&=K_{rr}\pm\frac{f_{rrr}}{\sqrt{\gamma_{rr}}}\,,\\
    U^\pm_\theta&=K_{\theta\theta}\pm\frac{f_{r\theta\theta}}{\sqrt{\gamma_{rr}}}\,,\\
    U^\pm_\phi&=\Pi\pm\frac{\psi_r}{\sqrt{\gamma_{rr}}}\,.
\end{align}

\subsection{Gauge Choices in the Bulk}\label{sec:gauge}

In this work, we will consider one particular gauge choice, that is that $\tilde\alpha$ and $\beta^r$ are time independent:
\begin{equation}
    \partial_t \tilde{\alpha} = \partial_t \beta^r =0. \label{Eq.gauge}
\end{equation}
Other types of gauge choices have been discussed in \cite{EC_Formulation}, but it is not clear to us that many of these choices do not change the characteristic structure of the evolution equations. If one considers a hyperbolic time evolution of the gauge variables, in general this will directly affect the characteristic structure, but it is not clear whether the characteristics are well-defined if $\tilde\alpha$ and $\beta^r$ are prescribed with elliptic gauge conditions. Therefore, we restrict our analysis to our gauge choice Eq.~(\ref{Eq.gauge}), where it is clear that the characteristic structure is as presented in the previous section, and leave other gauge choices for future work. 

\section{Boundary Conditions}\label{Sec:BCs}

We now restrict the problem to a spatial domain where $r\in[a,b]$. One way to gain some intuition on BCs for the scalar wave equation is to study the case of a static background. In such a case we have a time-like Killing vector $\xi^\nu$ and thus a conserved current $J^\mu=\xi^\nu T^\mu_{\ \nu}$. We then define the static energy
\begin{align}
    E_s=\int_a^b\left(\alpha\rho-\beta^rS_r\right)\sqrt{\gamma}\,dr\,,
\end{align}
and rewrite the conservation law $\nabla_\mu J^\mu=0$ in terms of the characteristics
\begin{align}
    \partial_t E_s=\frac{1}{4}\sqrt{\gamma_{rr}}\sqrt{\gamma}\left[(c_-U^-_\phi)^2-(c_+U^+_\phi)^2\right]\Big|_a^b\,\,.\label{Eq:Static_Energy}
\end{align}
The flux term on the right hand side suggests the BCs that can be applied to allow $E_s$ to enter/leave the domain:
\begin{align}
    U^+_\phi &= -k_a\frac{c_-}{c_+}U^-_\phi~~{\rm at}~~ r=a\,,\label{Eq:Static-BCa}\\ U^-_\phi &= -k_b\frac{c_+}{c_-}U^+_\phi~~{\rm at}~~ r=b \,,\label{Eq:Static-BCb}
\end{align}
for some coefficients $k_{a,b}$. While $k_{a,b}$ do not necessarily need to be constants, the values $k_{a,b}=\pm1$ keep $E_s$ exactly conserved in the domain for each boundary. 
In this work, a conservative BC that reverses the sign of an incoming pulse is referred to as a Dirichlet type (${k_{a,b}=-1}$), while one that retains the sign is referred to as a Neumann type (${k_{a,b}=+1}$). The motivation for this naming scheme comes from the fact that these BCs reduce to the classical Dirichlet ($\partial_t\phi=0$) and Neumann ($\partial_r\phi=0$) BCs in Minkowski space where $c_\pm=\pm 1$. It is important to note that we are considering domains and gauge conditions where  $c_+c_-<0$ throughout so that both boundaries always require BCs.

Once the metric becomes time dependent, $J^\mu$ is no longer conserved. Instead, the notion of what a reflection means needs to be rooted in a quantity that describes energy in a dynamical spacetime. One way of defining this is with the Misner-Sharp mass, which in spherical symmetry can be written as \cite{EC_Formulation}
\begin{align}
    M(r)=\frac{\sqrt{\gamma_{\theta\theta}}}{2}\left(1+\frac{U^+_\theta U^-_\theta}{\gamma_{\theta\theta}}\right)\,.\label{Eq:MS_mass}
\end{align}
To obtain the BCs on the incoming scalar modes $U^\pm_\phi$, we express the derivatives of the Misner-Sharp mass in terms of the state vector:
\begin{align}
    \partial_tM=&\ 4\pi\sqrt{\gamma_{\theta\theta}}\left[f_{r\theta\theta}(\beta^r\rho-\alpha S_r/\gamma_{rr})\right.\nonumber\\&\left.+K_{\theta\theta}(\alpha\rho+\alpha T-2\alpha S_{\theta\theta}/\gamma_{\theta\theta}-\beta^rS_r)\right]\,, \label{Eq:dtMS_mass}\\[0.2cm] \partial_rM= &\ 4\pi\sqrt{\gamma_{\theta\theta}}(f_{r\theta\theta}\rho-K_{\theta\theta}S_r)\,.\label{Eq:drMS_mass}
\end{align}
These relationships can be derived by taking the appropriate derivative of Eq.~(\ref{Eq:MS_mass}) and substituting time derivatives with the evolution equations and spatial derivatives with the constraints. It is interesting that these relationships can be expressed solely in terms of the state vector, which enables us to impose BCs on the scalar characteristics to control $M$.

To see this, let us integrate Eq.~(\ref{Eq:drMS_mass}) to define the Misner-Sharp ``energy'' enclosed within our domain
\begin{align}
    E_M=\int_a^b\frac{f_{r\theta\theta}\rho-K_{\theta\theta}S_r}{\sqrt{\gamma_{rr}\gamma_{\theta\theta}}}\sqrt{\gamma}\,dr\,.\label{Eq:MS_Energy}
\end{align}
In the continuum, we clearly have $E_M=M(b)-M(a)$. We can then write a conservation law by taking a time derivative and substituting Eq.~(\ref{Eq:dtMS_mass}). Written here for a massless scalar (i.e., $m=0$), we have
\begin{align}
    \partial_tE_M=\frac{\sqrt{\gamma}}{4\sqrt{\gamma_{\theta\theta}}}\left[c_+U^-_\theta(U^+_\phi)^2-c_-U^+_\theta(U^-_\phi)^2\right]\bigg|_a^b\,.\label{Eq:Energy}
\end{align}
The flux term on the right hand side suggests the general
BCs that can be applied to allow $E_M $ to enter/leave the domain:
\begin{align}
    U^+_\phi &= k_a\sqrt{\frac{c_-U^+_\theta}{c_+U^-_\theta}}\,U^-_\phi~~{\rm at}~~ r=a\,,\label{Eq:ScalarBCa}\\ U^-_\phi &= k_b\sqrt{\frac{c_+U^-_\theta}{c_-U^+_\theta}}\,U^+_\phi~~{\rm at}~~ r=b\,,\label{Eq:ScalarBCb}
\end{align}
where the coefficient values $k_{a,b}=\mp1$ define the Dirichlet/Neumann type BCs on the scalar characteristics that keep $E_M$ exactly conserved at each boundary. Interestingly, in the case of  $\partial_t\gamma_{\theta\theta}=0$, these reduce to the static background BCs, Eqns.~(\ref{Eq:Static-BCa}-\ref{Eq:Static-BCb}), which motivates their connection to the classical Dirichlet/Neumann BCs.

The expansion of outgoing null geodesics is proportional to the value of $U^-_\theta$, and thus the apparent horizon is the outermost surface where $U^-_\theta=0$. We call the surface where $c_+=0$ the characteristic horizon. Only when $c_+<0$ are BCs no longer required at $r=a$ as both characteristics are then leaving the domain. If $c_+c_-<0$ and $-U^-_\theta U^+_\theta<0$, i.e. the boundary $r=a$ is between the apparent horizon and the characteristic horizon, the BCs at $r=a$ are not well-defined unless $k_a=0$. We will not consider any simulations where such cases are present in this work.

The BCs on the incoming angular characteristics $U_\theta^\pm$ are given by Eqn.~(\ref{Eq:MS_mass}):
\begin{align}
    U^+_\theta&=\frac{2M(a)\sqrt{\gamma_{\theta\theta}}-\gamma_{\theta\theta}}{U^-_\theta}~~{\rm at}~~ r=a\,,\label{Eq:AngularBCa}\\ U^-_\theta&=\frac{2M(b)\sqrt{\gamma_{\theta\theta}}-\gamma_{\theta\theta}}{U^+_\theta}~~{\rm at}~~ r=b\,,\label{Eq:AngularBCb}
\end{align}
where the values of $M(a)$ and $M(b)$ at each time step are given by integrating their time derivatives with the scalar BCs applied
\begin{align}
    \partial_tM(a) &= -\pi\tau_a c_-\sqrt{\gamma_{rr}\gamma_{\theta\theta}}U^+_\theta (U^-_\phi)^2~~{\rm at}~~ r=a\,,\\
    \partial_tM(b) &= \ \ \, \pi \tau_b c_+\sqrt{\gamma_{rr}\gamma_{\theta\theta}}U^-_\theta (U^+_\phi)^2~~{\rm at}~~ r=b\,,
\end{align}
where $\tau_{a,b}=(1-k_{a,b}^2)$. We thus see that, in the case of Dirichlet/Neumann type BCs, the Misner-Sharp mass at both boundaries remains constant. 

As long as $\beta^r>0$, the boundary $r=b$ has three additional incoming modes $U^0_r$, $U^0_\theta$, and $U^0_\phi$. As suggested in \cite{EC_Boundaries}, these can all be fixed using the constraints (\ref{Eq.fr}-\ref{Eq:C_phi}) by replacing the $r$ derivatives in their evolution equations:
\begin{align}
\partial_t \phi= &\ \beta^r(\psi_r) -\alpha\Pi\,,\\
\partial_t\gamma_{rr} = &\  \beta^r\left(2f_{rrr}-8\frac{\gamma_{rr}f_{r\theta\theta}}{\gamma_{\theta\theta}}\right) \nonumber\\&+ 2\gamma_{rr}\partial_r\beta^r - 2\alpha K_{rr}\,, \\
\partial_t\gamma_{\theta\theta} = &\  \beta^r(2f_{r\theta\theta})  -2\alpha K_{\theta\theta}\,,
\end{align}
all evaluated at $r=b$.

While fixing the masses $M(a)$ and $M(b)$, and scalar characteristics (Eqs.~\ref{Eq:ScalarBCa}-\ref{Eq:ScalarBCb}), fixes all the physical degrees of freedom at the boundaries, the incoming radial characteristics $U_r^\pm$ are left arbitrary, and thus should be connected to the residual gauge freedoms. Although we have specified the gauge dynamics in the bulk by choosing $\tilde\alpha$ and $\beta^r$, additional gauge freedom can propagate into the domain from outside, and thus needs to be specified by fixing the incoming radial modes.

We shall consider a few choices for the BCs of $U_r^\pm$. First, one could impose $\partial_tU^\pm_r=0$ at the corresponding boundaries: 
\begin{align}
U^+_r(t)&=U^+_r(0)~~{\rm at}~~ r=a\,,\\ U^-_r(t)&=U^-_r(0)~~{\rm at}~~ r=b\,,
\end{align}
which ensures there are no incoming radial characteristics into the domain.
Next, we could impose $\partial_tK_{rr}=0$:
\begin{align}
U^+_r(t)&=-U^-_r(t)+2K_{rr}(0)~~{\rm at}~~ r=a\,,\label{Eq:RadialBCa}\\ U^-_r(t)&=-U^+_r(t)+2K_{rr}(0)~~{\rm at}~~ r=b\,,\label{Eq:RadialBCb}
\end{align}
and finally we could impose $\partial_tf_{rrr}=0$:
\begin{align}
U^+_r(t)&=U^-_r(t) -\frac{2f_{rrr}(0)}{\sqrt{\gamma_{rr}(t)}}~~{\rm at}~~ r=a\,,\\ U^-_r(t)&=U^+_r(t) +\frac{2f_{rrr}(0)}{\sqrt{\gamma_{rr}(t)}}~~{\rm at}~~ r=b\,,
\end{align}
It is not clear if different choices may be more advantageous, but we find that for long term stability, the condition $\partial_tK_{rr}=0$ at both boundaries seems to work the best.

\section{Initial Conditions}\label{Sec:ICs}

For the initial background, we consider a Schwarzschild black hole in Kerr-Schild coordinates, but with a $r$-dependent mass function:
\begin{align}
    \alpha^{-2} = 1+\frac{2M(r,0)}{r}\,&,\quad
    \beta^r =\frac{2M(r,0)}{r}\alpha^2\,,\nonumber\\\quad
    \gamma_{rr} = \alpha^{-2}\,&,\quad
    \gamma_{\theta\theta} = r^2\,,
\end{align}
The initial values of $f_{rrr}$ and $f_{r\theta\theta}$ are formed from constraint Eqs.~(\ref{Eq.fr}) and (\ref{Eq.f_theta}). The initial values of extrinsic curvature components are formed via Eqs.~(\ref{Eq.EC_r}) and (\ref{Eq.EC_theta}), with $\partial_t\gamma_{\theta\theta}=0$ assumed, but $\partial_t\gamma_{rr}$ kept arbitrary. The yet unspecified functions $M(r,0)$ and $\partial_t\gamma_{rr}$ will be used to satisfy the constraints. For the scalar field, a spherical pulse of amplitude $A$, total width $2\sigma$, and location $\mu$ is modeled as a section of a polynomial:
\begin{align}
    \phi(r,0) =\frac{A}{r\sigma^8}[r-(\mu-\sigma)]^4[r-(\mu+\sigma)]^4\,,\label{Eq:scalar_init}
\end{align}
and is set to zero outside of $(\mu-\sigma)\leq r\leq(\mu+\sigma)$. The degree of this polynomial assures that the first three derivatives are continuous. The initial condition on $\psi_r$ is obtained by taking the $r$ derivative. For the initial condition on $\Pi$, a common choice is $\Pi=0$, which describes a pulse that will break into two parts traveling at speeds $c_+$ and $c_-$. Another choice is to specify that the pulse is initially traveling at speed $c_-$, such that the initial time derivative is
\begin{align}
    \partial_t\phi(r,0) = -\frac{8Ac_-}{r\sigma^8}\left[(r-\mu)^2-\sigma^2\right]^3(r-\mu)\,,\label{Eq:IngoingIC}
\end{align}
and is zero outside of $(\mu-\sigma)\leq r\leq(\mu+\sigma)$. The initial value of $\Pi$ is then completed with the definition ${\Pi=-(\partial_t\phi-\beta^r\psi_r)/\alpha}$. 
Given initial conditions on the scalar field, the Hamiltonian and momentum constraints must be satisfied at the initial time. With the two functions $M(r,0)$ and $\partial_t\gamma_{rr}$, one can show that these constraints are satisfied if 
\begin{align}
    \partial_rM(r,0)&=4\pi r^2\left(\alpha\rho-\beta^rS_r\right)/\alpha\,,\label{Eq:InitialMass}\\
    \partial_t\gamma_{rr}&=-8\pi r S_r/\alpha\,.
\end{align}
These are easy conditions to satisfy, as $M(r,0)$ can be integrated with a standard solver starting at $r=a$ with initial internal mass $M_0$ or from $r=b$ with initial total mass $M_\mathrm{tot}$. The value of $\partial_t\gamma_{rr}$ can be replaced outright in the initial value of $K_{rr}$.


\section{Numerical Implementation}\label{Sec:Numerics}

In this section, we will introduce our numerical implementation of the IBVPs we are considering. We use Summation-By-Parts (SBP) derivative operators in an effort to preserve numerical stability by ensuring the energy of the system is bounded. In order to impose boundary conditions (BCs) in a stable fashion we use simultaneous approximation terms (SATs). Finally, we use numerical dissipation to prevent instability due to high-frequency noise. 

\subsection{Summation-By-Parts Derivative Operators}\label{Sec:SBP}

A popular way to maintain stability in numerical hyperbolic IBVPs is to use finite differencing operators that satisfy SBP, which is the discrete analog of integration by parts. Using SBP operators can be proven to be closely related to the well-posedness and the numerical stability of many conservative problems \cite{SBP_Review,Shifted_Wave}. In order to introduce our covariant SBP scheme, we review some covariant integration properties. In a 3-dimensional domain $V$ with coordinates $x$ and the 2-dimensional boundary of that domain $\partial V$ with coordinates $\xi$, the covariant Gauss's theorem states that for a vector field $F^i$
\begin{align}
     \int_V \nabla_i F^i \sqrt{\gamma}\, d^3x = \oint_{\partial V} n_iF^i\sqrt{s}\, d^2\xi\,,
\end{align}
where the determinants of the metric on $V$ and $\partial V$ are $\gamma$ and $s$ respectively and $n_i$ is the unit normal to the boundary $\partial V$. If the vector field is represented by a scalar factor and a vector factor as $F^i=u v^i$, in spherical symmetry, this property can be written as
\begin{align}
     \int^b_a u\nabla_r v^r \sqrt{\gamma}\, dr+\int^b_a (\partial_r u)v^r \sqrt{\gamma}\, dr = uv^r\sqrt{\gamma}\Big|_a^b\,,\label{Eq:CovariantIntegration}
\end{align}
where $\nabla_rv^r=(\sqrt{\gamma})^{-1}\partial_r(\sqrt{\gamma}\,v^r)$ is the covariant divergence. This can be thought of as the covariant version of integration by parts in 1-dimension. Our goal will be to develop a numerical implementation of our IBVP that obeys a discrete analog of this property in an effort to maintain numerical stability.

For our spatial domain $r\in[a,b]$ we define a grid where all functions will be sampled with $n$ points spaced by a distance $h$. An $n\times n$ finite differencing operator $D$ paired with a symmetric and positive definite $n\times n$ norm operator $\Sigma$ are said to satisfy SBP if \cite{SBP_Review}
\begin{align}
     \Sigma D + D^{\rm T}\Sigma = B\,,
     \label{Eq:SBP}
\end{align}
where the boundary operator $B=\mathrm{diag}(-1,0,\dots,0,1)$. For this property to mimic the covariant property Eq.~(\ref{Eq:CovariantIntegration}), we insert a matrix $\Gamma$
with the values of $\sqrt{\gamma}$ injected along the diagonal:
\begin{align}
     W \nabla + D^{\rm T}W = B\Gamma\,.
     \label{Eq:SBPcov}
\end{align}
The operator $\nabla \equiv \Gamma^{-1}D\Gamma$ approximates the covariant 3-divergence of 3-vectors, the operator $D$ approximates the covariant scalar gradient, and $W\equiv\Sigma\Gamma$ is the covariant norm operator and approximates covariant integration. Then, in one spatial dimension, for a scalar grid function $u$ and the one non-zero component of a 3-vector grid function $v^r$, 
\begin{align}
u^\mathrm{T}W \nabla v^r + (D u)^\mathrm{T}W v^r  = u v^r\sqrt{\gamma}\Big|_a^b\,,\label{Eq:General_SBP}
\end{align}
which directly mimics the continuous property Eq.~(\ref{Eq:CovariantIntegration}).

In the case of the wave equation on a static background, one can write the energy conservation law in a directly analogous way to the covariant property Eq.~(\ref{Eq:CovariantIntegration}) (see Appendix \ref{App:SBP}), and thus when the system is discretized using the SBP operators, the system remains strictly stable in the sense that the energy is bounded \cite{SBP_Review,Shifted_Wave}. It is no longer clear how to properly do this when the spacetime becomes dynamic. The state vector in the EC system is defined in terms of lower-index functions ($f_{ijk}$), which runs counter to the covariant SBP property that is defined in terms of an upper-index function. We have discovered one way that seems to be stable in at least all of the cases considered here, but a more elegant discretization paradigm may exist, perhaps one where the SBP properties can be made to mimic the dynamic conservation law Eq.~(\ref{Eq:Energy}), and the equations of motion can be written in terms of covariant divergences of upper-index functions. In the numerical scheme presented here, the evolution equations written in Section~\ref{Sec:EC_System} are discretized by replacing scalar gradients with an SBP operator (i.e. $\partial_r\rightarrow D$) and where present the operator $\mathcal{D}_r\rightarrow \Gamma^{-1}D\Gamma$. 

In this work, we use the $D_{6-3}$ SBP operator defined in \cite{SBP_2007} that minimizes the so-called average boundary truncation error. This operator is 6\textsuperscript{th} order accurate in the interior and 3\textsuperscript{rd} order accurate near the boundaries. It was reported in \cite{SBP_2007} that this operator achieves near 4\textsuperscript{th} order global convergence with a diagonal norm operator, which in principle will pair nicely with standard 4\textsuperscript{th} order Runge-Kutta time integration.

\begin{figure*}[t]
\includegraphics[width=\textwidth]{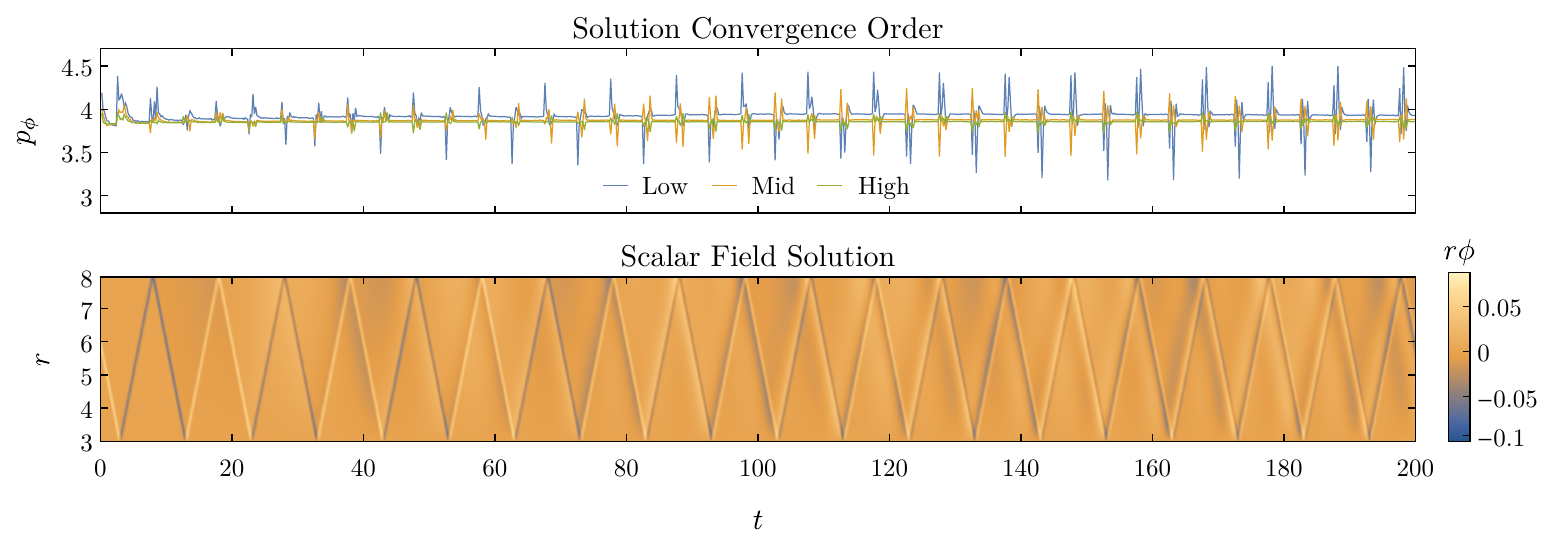}
\caption{Convergence and solution for the flat spherical wave equation. At the top, the solution convergence order settles to about 4\textsuperscript{th} order even at coarse resolutions, and deviates only when a reflection occurs where the error is dominated by the boundary region accurate to 3\textsuperscript{rd} order. At the bottom we see the wave solution in the $r$--$t$ plane, where the pulse travels in straight lines since $c_\pm=\pm 1$. }\label{Fig:SphericalBox}
\end{figure*}

\subsection{Application of Boundary Conditions}

The BCs are applied using SATs. This method applies BCs in a weak fashion, adding an exponential decay term to the evolution equations at the boundary. For two arbitrary characteristic grid functions $U^\pm(r,t)$ with characteristic speeds $c_\pm$, the application of the general BCs $U^+(a,t)=g_a(t)$ and $U^-(b,t)=g_b(t)$ is
\begin{align}
    \partial_t U^+(t)&= \cdots + \frac{s_a}{\Sigma_{11}}\left[g_a(t)-U^+(t)\right]~~{\rm at}~~ r=a\,,\\\partial_t U^-(t)&= \cdots + \frac{s_b}{\Sigma_{nn}}\left[g_b(t)-U^-(t)\right]~~{\rm at}~~ r=b\,,
\end{align}
with strengths $s_a$ and $s_b$ which dictate the exponential decay scale \cite{SBP_Review}. One can show that for proper SBP energy conservation, in the context of the wave equation on a static background, the strengths must be equal to the magnitude of the incoming characteristic speeds \cite{Shifted_Wave}. At the left boundary we have $s_a\rightarrow c_+$ and at the right boundary we have $s_b\rightarrow -c_-$. In the case of the wave equation around a static background black hole in Kerr-Schild coordinates, we have $s_b=1$ and $0<s_a<1$ where $s_a=0$ when $a=2M_0$ as the boundary is then placed exactly on the horizon, where BCs are not to be imposed anyway since there are no longer any incoming characteristics into the domain at $r=a$.



\subsection{Numerical Dissipation}

To stabilize nonlinear evolution it is common to add numerical dissipation, ensuring that high frequency noise does not destabilize the evolution. We use the $A_6$ numerical dissipation operator that applies to the $D_{6-3}$ operator defined in \cite{SBP_2007}, and it is added to the right hand side of the evolution equations as
\begin{align}
    \partial_t\vec U=\cdots + \frac{\epsilon}{2^6}A_6 \vec U\,.
\end{align}
where $\epsilon>0$ is the amount of dissipation, which is usually chosen to be of order unity. The coefficients of both operators $D_{6-3}$ and $A_6$ can be found in \cite{SBP_2007}.

\section{Results}\label{Sec:Results}

First, to demonstrate long term stability and energy conservation of this framework, an IBVP with two distinct reflection boundaries, one of Dirichlet type and one of Neumann type, will be demonstrated in several situations. Then we will investigate situations where the boundary $r=a$ is brought very close to the horizon of the black hole. Finally, we discuss the movement of the apparent horizon and situations where the initial pulse can reflect without the apparent horizon moving into the domain.

\begin{figure*}[t]
\includegraphics[width=\textwidth]{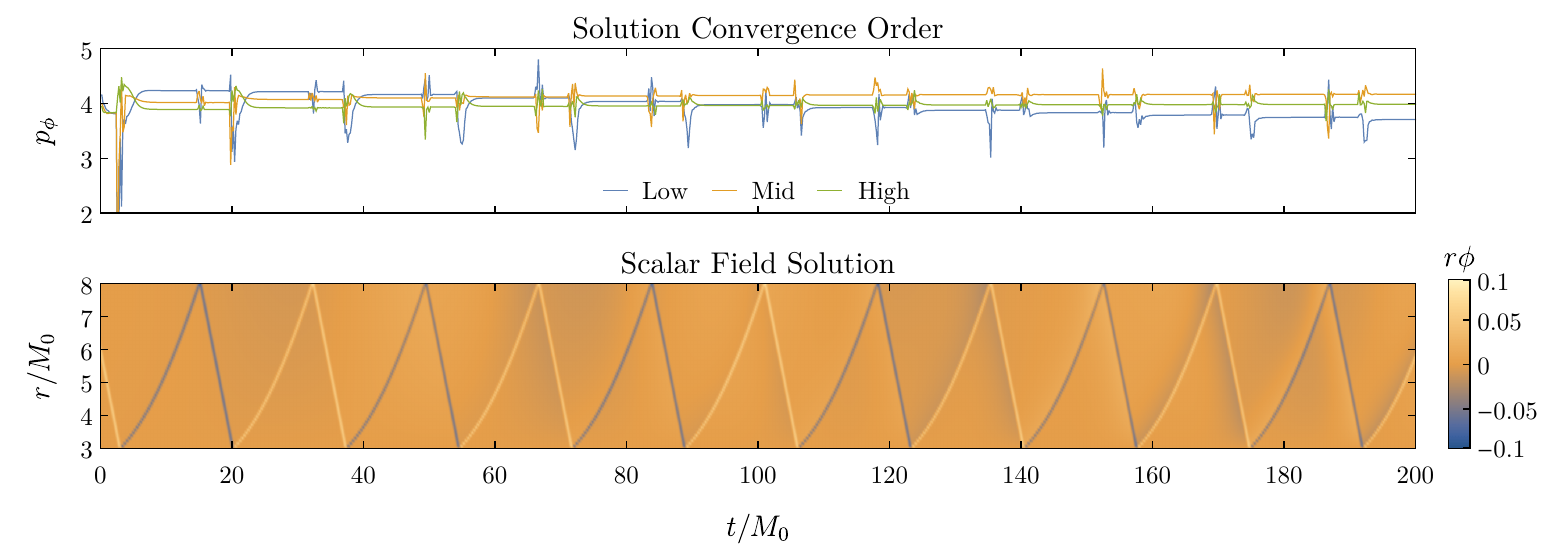}
\caption{Convergence and solution for the static Schwarzschild background wave equation in Kerr-Schild coordinates. At the top, the solution convergence order settles to about 4\textsuperscript{th} order as the resolution is increased, deviating when a reflection occurs where the error is dominated by the boundary region accurate to 3\textsuperscript{rd} order. At the bottom we see the wave solution in the $r$--$t$ plane, where it is apparent that a pulse propagating in the $-r$ direction has a constant characteristic speed while one traveling in the $+r$ direction appears to accelerate along curves.}\label{Fig:KerrSchild}
\end{figure*}

\subsection{Enclosed Reflecting Boundary Problems}

Here we consider several enclosed reflecting boundary problems. These problems involve an initial pulse and BCs of either Dirichlet or Neumann type so that energy is conserved in the continuum limit. We consider three situations for this type of problem, first a Minkowski spherical wave equation, second a static background Schwarzschild black hole in Kerr-Schild coordinates, and third the fully dynamic problem around a black hole. In all of these examples, the initial conditions and evolutions are kept as similar as possible to aid in a direct comparison.

Here, we consider a spatial domain $r\in[3,8]M_0$ and a temporal domain $t\in[0,200]M_0$. As the mass $M_0$ is not relevant in Minkowski space, we simply take $r\in[3,8]$ and $t\in[0,200]$ in this case. We use a Dirichlet type BC for $r=a$ (i.e. $k_a=-1$) and a Neumann type BC for $r=b$ (i.e. $k_b=1$). For the problems with a black hole, the internal mass is set to $M_0=1$. The initial pulse follows the model Eq.~(\ref{Eq:scalar_init}) with parameters $\mu=6M_0$, $2\sigma=M_0$, and $A=0.05/M_0$. We use initial condition Eq.~(\ref{Eq:IngoingIC}) so that the pulse is initially traveling with speed $c_-=-1$. 

The time integration is done using 4\textsuperscript{th} order Runge-Kutta time stepping, with time step size $h_t=h/4$. We also apply the numerical dissipation operator with strength $\epsilon=1$ in all cases so that the energy loss is directly comparable to the fully dynamic case.


\begin{figure*}[t]
\includegraphics[width=\textwidth]{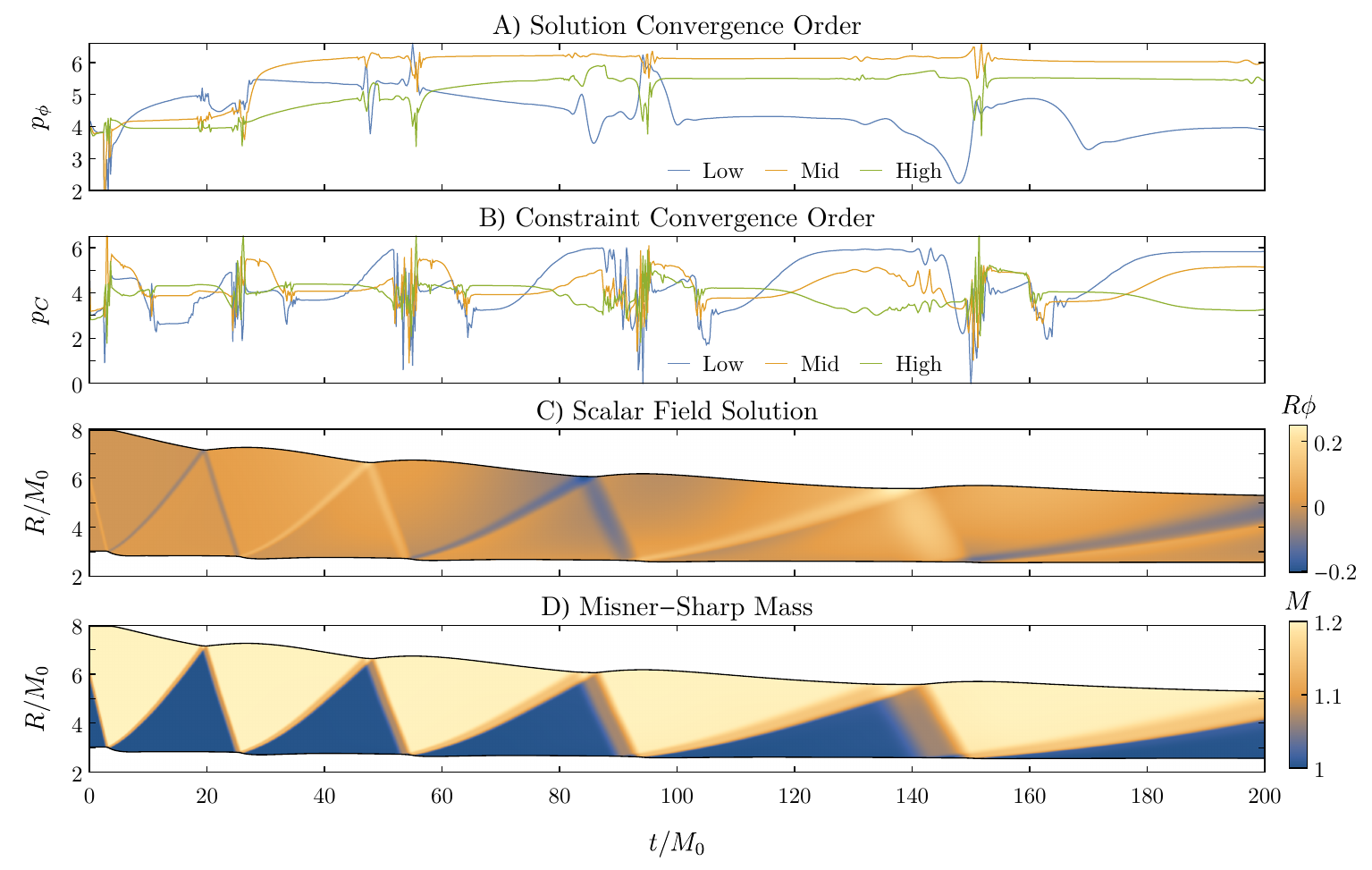}
\caption{Convergence and solution for the fully dynamic IBVP. Plot A) shows the solution convergence order that starts near 4\textsuperscript{th} order but changes dramatically as the solution becomes more dependent on boundary interactions and the solution errors move out of phase with each other. Plot B) shows the constraint convergence order slowly settles to about 4\textsuperscript{th} order as the resolution is increased. Plot C) shows the scalar field solution shown as a function of the areal radius $R=\sqrt{\gamma_{\theta\theta}}$ over coordinate time $t$. It is apparent that over the course of the evolution, the inner boundary falls toward the horizon by $\sim 0.5M_0$ and the outer boundary by $\sim 2.7M_0$. The pulse also appears to disperse much more than in Fig. \ref{Fig:KerrSchild}, due to the variation of $c_\pm$ as the Misner-Sharp mass $M(r)$ changes across the pulse width. Plot D) shows the Misner-Sharp mass in the same $R$--$t$ plane, demonstrating the coupling to gravity.}\label{Fig:Dynamic}
\end{figure*}

\begin{figure}[h]
\includegraphics[width=0.45\textwidth]{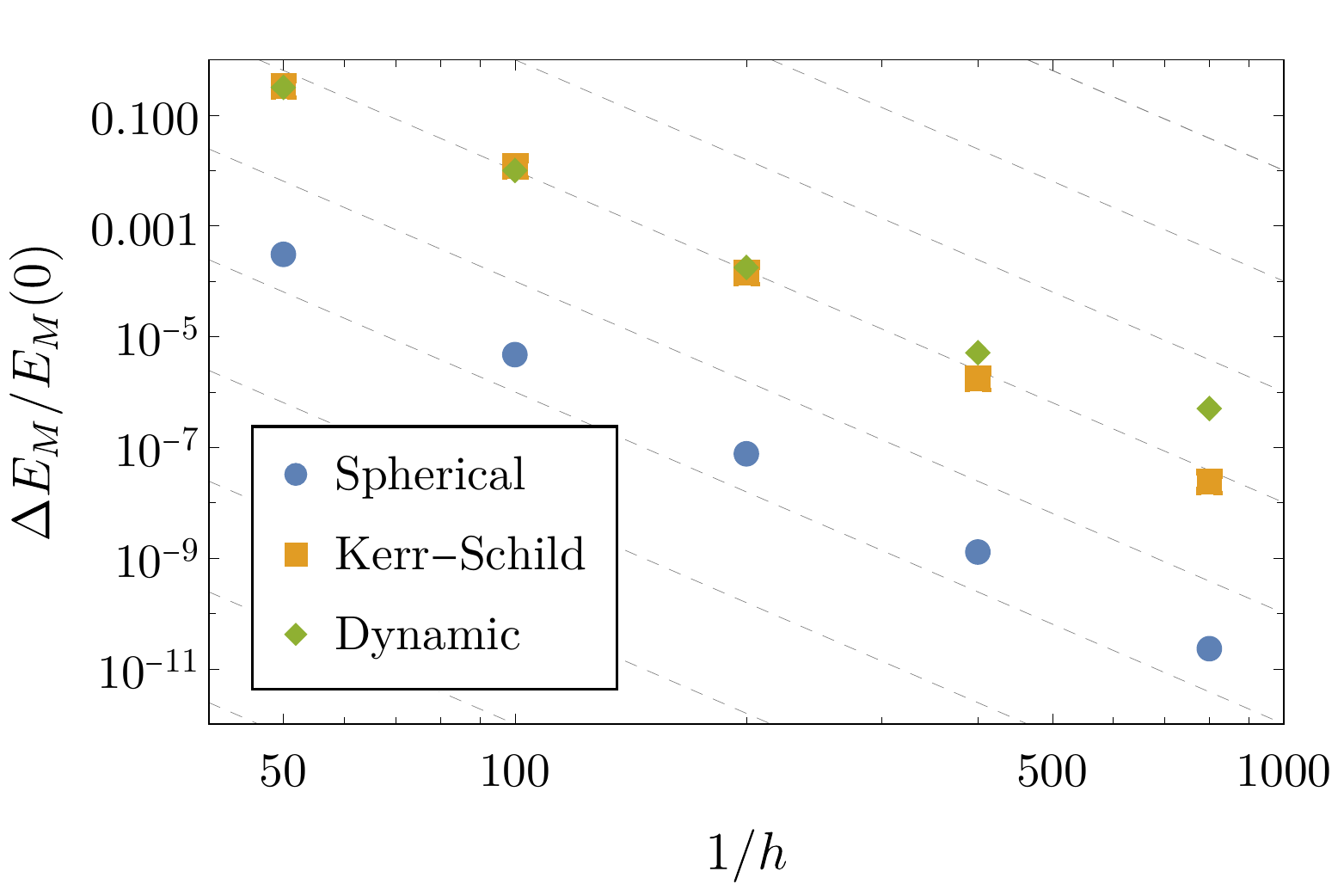}
\caption{Energy loss for various resolutions of the three enclosed boundary scenarios. Here, $\Delta E_M=E_M(0)-E_M(200)$ and dashed lines are $\propto h^{6}$. The static background cases converge to exact conservation at 6\textsuperscript{th} order. The dynamic case appears to diverge from this power law at the highest resolutions.}\label{Fig:Energy}
\end{figure}

\subsubsection{Static Minkowski Background}

Figure~\ref{Fig:SphericalBox} depicts the IBVP of the flat spherically symmetric wave equation. To demonstrate the convergence of the solution, we define the solution convergence order:
\begin{align}
   p^{(i)}_\phi(t)=\log_2\left(\,\frac{||\phi^{(i-2)}(t)-\phi^{(i-1)}(t)||}{||\phi^{(i-1)}(t)-\phi^{(i)}(t)||}\,\right)
\end{align}
where the $\phi^{(i)}$ are the solutions obtained with a certain grid spacing $h^{(i)}$ and where the $\ell_2$ norm is defined for a grid function $f$ as
\begin{align}
   ||f||^2 &= \int^b_af^2\sqrt{\gamma}\,dr\,,\nonumber\\||f||^2&\approx ||f^{(i)}||^2=\big(f^{(i)}\big)^{\rm T}Wf^{(i)}\,.
\end{align}
We use $h^{(i)} = h^{(i-1)}/2$ as the resolution is doubled each run. Five runs were conducted starting at $h^{(1)}=0.02M_0$. This leaves three subsets of resolutions to calculate the solution convergence order $p_\phi^{(i)}$, referred to as Low ($i=3$), Mid ($i=4$), and High ($i=5$). The $D_{6-3}$ SBP operator coupled with 4\textsuperscript{th} order Runge-Kutta time stepping demonstrates very consistent 4\textsuperscript{th} order convergence in this case. Since the characteristic speeds here are $c_\pm=\pm1$, the pulse follows a straight line in the $r$--$t$ plane and returns back to $r=6$ at precisely $t=200$. 


\subsubsection{Static Schwarzschild Background}

Figure~\ref{Fig:KerrSchild} depicts the numerical solution to the IBVP with a static Schwarzschild background in Kerr-Schild coordinates. The same five resolutions are used to obtain the Low, Mid, and High convergence order plots, which shows once again consistent 4\textsuperscript{th} order convergence as the resolution is increased. Since the characteristic speeds here are $c_-=-1$ and $c_+=(r-2M_0)/(r+2M_0)$, pulses traveling in the $-r$ direction follow straight lines in the $r$--$t$ plane and stay the same width, but pulses traveling in the $+r$ direction appear to accelerate along curves and have a characteristic speed that varies across the pulse, so the width increases as the pulse propagates in the $+r$ direction.

\subsubsection{Dynamic Schwarzschild Background}

Here we demonstrate a fully dynamic IBVP, where the scalar pulse is coupled to gravity. Along with the solution convergence order, we also define the constraint convergence order
\begin{align}
   p^{(k)}_C=\log_2\left(\,\frac{||C_{\rm tot}^{(k-1)}||}{||C_{\rm tot}^{(k)}||}\,\right)\,,
\end{align}
where the norm of the constraints is defined as
\begin{align}
  ||C_{\rm tot}||^2 &\equiv  \int^b_a(C^2+\gamma^{ij}C_iC_j)\sqrt{\gamma}\,dr\,\\
   ||C_{\rm tot}||^2&\approx ||C^{(k)}_{\rm tot}||^2 \nonumber\\&\equiv \big(C^{(k)}\big)^\mathrm{T}WC^{(k)} + \big(C_r^{(k)}\big)^{\mathrm{T}}W\big(C^r\big)^{(k)} 
\end{align}
This somewhat simpler definition for the constraint convergence order is possible because we know the exact solution (i.e. $C=C_r=0$) whereas we don't necessarily  know the exact solution for the scalar field $\phi$.

We use total reflection BCs on the scalar field Eqns.~(\ref{Eq:ScalarBCa}) and (\ref{Eq:ScalarBCb}) with $k_a=-1$ and $k_b=+1$. We use angular BCs Eqns.~(\ref{Eq:AngularBCa}) and (\ref{Eq:AngularBCb}) with $M(a)=M_0=1$ and $M(b)=M_\mathrm{tot}$ for all time, where the value of $M_\mathrm{tot}$ is given from the integration of Eq.~(\ref{Eq:InitialMass}) at the initial time slice and in this case is $M_\mathrm{tot}\sim 1.18M_0$. And finally, we use radial BCs Eqns.~(\ref{Eq:RadialBCa})~and~(\ref{Eq:RadialBCb}).

The results are plotted in Figure \ref{Fig:Dynamic}. At the initial time slice, $c_+=[r-2M(r)]/[r+2M(r)]$, so the outgoing characteristic speed varies across the pulse as in the static background case, but also as $M(r)$ changes across the pulse width, so there is an additional dispersion due to the coupling to gravity. In this particular gauge, $\partial_tc_\pm=\pm\tilde\alpha\partial_t\gamma_{\theta\theta}$, so the characteristic speeds change proportional to the change in the areal radius, $R=\sqrt{\gamma_{\theta\theta}}$. We notice that our choice of BCs causes the boundaries to move inward in $R$ , making the reflected pulse at the inner boundary move more slowly as time progresses.

In the continuum limit, each of the above IBVPs conserves the energy Eq.~(\ref{Eq:Energy}). Figure \ref{Fig:Energy} shows that the energy loss at $t=200M_0$ in these discrete problems converges to zero at about 6\textsuperscript{th} order. Since the strength of the SAT boundary at $r=a$ depends on the characteristic speed $c_+$, it is applied more weakly in the Kerr-Schild and dynamic cases, leading to significantly more energy dissipation at the boundary. This is apparent in Figure~\ref{Fig:Energy} as the energy decreases about three orders of magnitude more in the Kerr-Schild case than the flat spherical case at $t=200M_0$, although still converges to zero at 6\textsuperscript{th} order.


\subsection{Reflections close to the horizon}

Black hole echoes are usually discussed in the context of quantum gravity, suggesting that quantum phenomena may result in a reflecting surface. If this is the case, one might expect such a surface to be a distance on the order of the Planck scale away from the horizon (e.g.,\cite{Cardoso:2016oxy,Abedi:2016hgu,Oshita:2019sat}). To simulate this, one might be interested in applying reflecting BCs very close to the horizon. Although this problem is well-posed in principle, in practice it can be difficult to achieve numerically.

\begin{figure}[ht]
\includegraphics[width=0.5\textwidth]{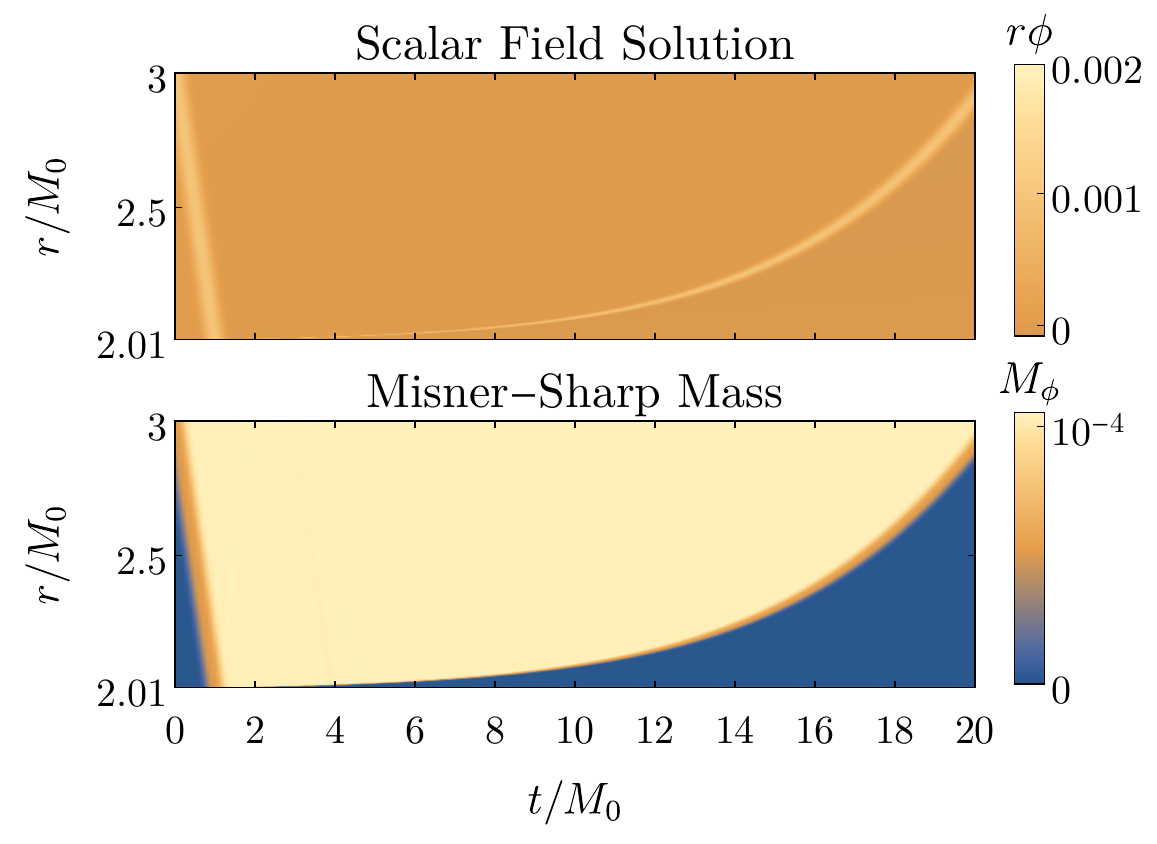}
\caption{A fully dynamic simulation of a scalar pulse incident on a reflection boundary at $a=2.01M_0$. Here the Misner-Sharp mass of the pulse is $M_\phi=M-M_0$.}\label{Fig:Close}
\end{figure}

In the context of the static black hole background, if a pulse of width $w$ is incident on a reflecting surface, the reflected pulse will have width $\sim w|c_-/c_+|$, so one should be sure that $h$ is at least several times smaller than this. There is also usually a significant amount of high frequency noise that propagates at high speed off of the boundary $r=a$, which may need to be controlled with numerical dissipation. When $c_+$ is very small at $r=a$, the SATs become very weak and thus lose much more energy, so in order for the discrete problem to conserve energy  to a certain tolerance, much more resolution might be required than the naive estimate $h<w|c_-/c_+|$.

Figure \ref{Fig:Close} depicts a fully dynamic simulation of a scalar pulse incident on a reflecting surface at coordinate radius $a=2.01M_0$ with $k_a=+1$. For a stellar mass black hole, $a=2.01M_0$ is of course far from Planck scale. We mean this as a demonstration of the difficulties one would face in simulating Planck-scale reflections. 
To achieve a conservation of the Misner-Sharp energy $E_M$ to within a tolerance of $\sim 10^{-5}M_0$, a grid spacing of $h\sim6\times10^{-5}M_0$ was needed. The parameters of the initial scalar pulse are $A=0.001/M_0$, $2\sigma=M_0$, and $\mu=3M_0$, which results in a total mass $M_\mathrm{tot}\sim (1+ 10^{-4} )M_0$ which seems to be about the highest mass allowed (given our gauge and BCs) before the apparent horizon moves into the domain and the boundary can no longer reflect.

\subsection{Apparent Horizons}

Finally, as a response to \cite{GuoMathur}, we can study when an apparent horizon moves into the domain of the simulation, making BC Eq.~(\ref{Eq:ScalarBCa}) ill-defined. In \cite{GuoMathur}, Guo and Mathur argue that for merging equal mass  black holes, one should never expect echoes of GWs due to the formation of an apparent horizon that envelopes the waves. They make the assumption that the waves have a wavelength on the same order of the black hole masses, but any localized wavepacket should contain shorter, as well as longer, wavelengths. If a black hole absorbs a sufficiently localized infalling wavepacket with a total mass $M_\phi=M_\mathrm{tot}-M_0$, then the apparent horizon should move from $R=2M_0$ to $R=2M_\mathrm{tot}$ as $t\rightarrow\infty$. This suggests that a reflecting surface should be located at $R>2M_\mathrm{tot}$ to ensure that the wavepacket does not become enveloped by a trapped surface. We demonstrate, however, that this estimate can be relaxed for sufficiently wide pulses, and thus the frequency content of a wavepacket plays an important role in whether a reflection can occur.

We consider a domain $r\in [2.1,203]M_0$ with the initial pulse location $\mu=103M_0$ and vary its width from $\sigma=10M_0$ to $\sigma=100M_0$. We consider two situations, one where the boundary $r=a$ has Neumann type reflective condition $k_a=+1$, and one where it has transmitting (or absorbing) condition $k_a=0$. If the mass of the initial pulse $M_\phi$ is too large, the apparent horizon will move into the domain from its initial position of $r=2M_0$ which is initially outside of the domain. We then maximize the mass of the initial pulse by adjusting the amplitude $A$ to just before this occurs, and plot the results of this maximum mass for the varying initial widths. The transmitting case $k_a=0$ demonstrates reflection just from the Schwarzchild potential, a feature that occurs regardless of BCs we apply. 

\begin{figure}[t]
\includegraphics[width=0.45\textwidth]{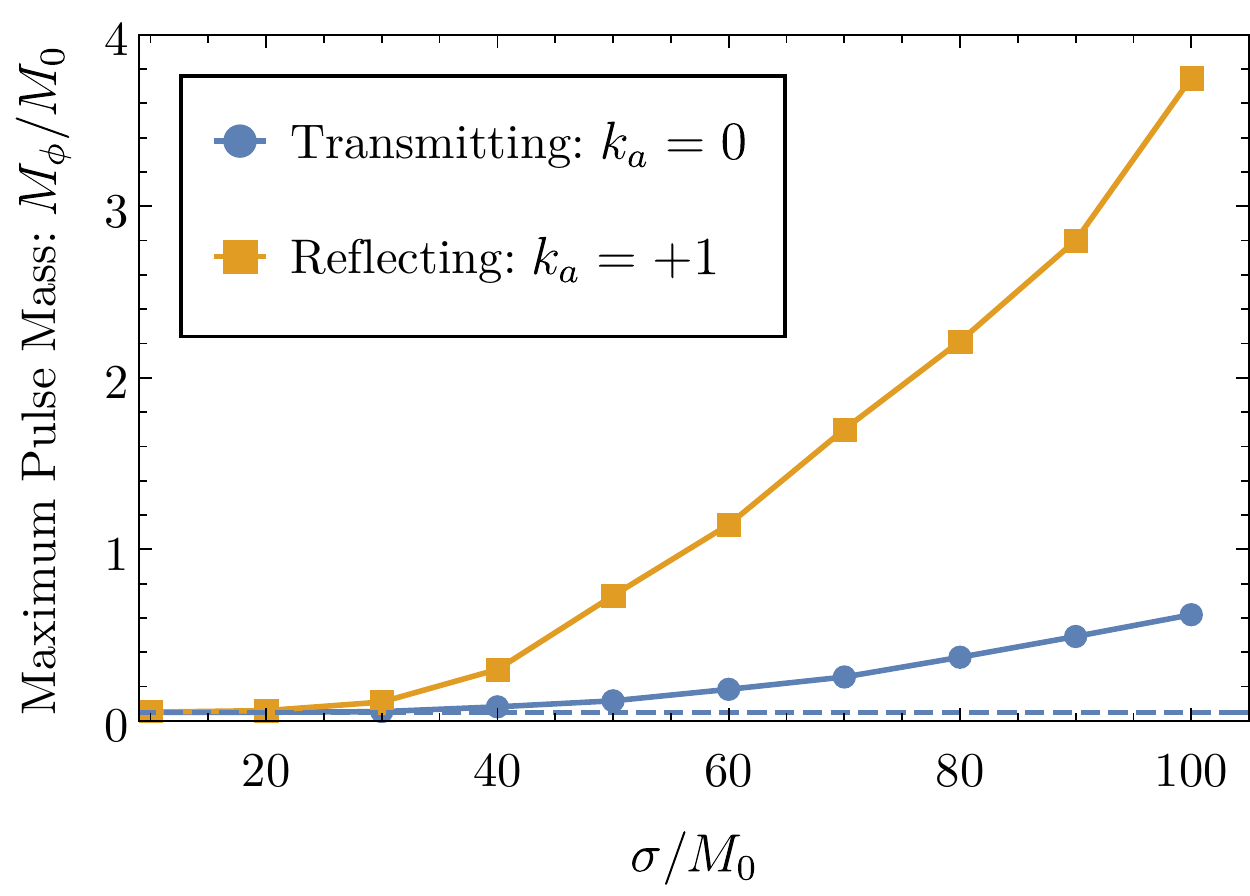}
\caption{Plot of the maximum pulse mass $M_\phi$ of an initial pulse before the apparent horizon moves into the domain for various widths of the initial pulse. Each case is asymptotic to $M_\phi=0.05M_0$ as the width goes to zero ($\sigma \to 0$), which corresponds to an infinitely sharp pulse with $a=2M_{\rm tot}$. }\label{Fig:ReflectionVsWidth}
\end{figure}

Figure~\ref{Fig:ReflectionVsWidth} shows that the case $k_a=+1$ (reflective BC) allows a significantly higher maximum mass $M_\phi$ than the transmitting case $k_a=0$ (absorptive BC), which suggests that there is a possibility for low frequency constituents of the incident scalar pulse to reflect near the horizon in a way discernible from reflection due just  to the Schwarzchild potential. Indeed, low frequency (or wide) pulses, comparable to the Hawking frequency $1/(8\pi M_0)$ are also expected to have the largest reflection (from the horizon) based on quantum mechanical arguments \cite{Oshita:2019sat}.

It is important to note that as we have observed previously, our gauge conditions tend to move the boundaries inward, and so the apparent horizon can move into the domain either because it is growing in areal radius, or due to the boundary $r=a$ shrinking in areal radius. The orange curve in Figure~\ref{Fig:ReflectionVsWidth} should then be thought of as a lower bound that could be higher if the boundary was fixed in areal radius (through a different choice of gauge/BCs).

\section{Conclusion and Future Prospects}

In this work, we developed and demonstrated a full initial boundary value problem in numerical relativity in spherical symmetry. In particular, we derived boundary conditions (BCs) based on conservation laws involving a quasi-local mass-energy measure, which is the main result of this work. These BCs are general, allowing one to dictate how much mass-energy enters or leaves the domain. We demonstrated how this framework can model black hole echoes, which has mostly been discussed previously either within linear perturbation theory that ignored backreaction or order of magnitude estimates.

However, we only considered a gauge condition where $\tilde\alpha$ and $\beta^r$ were both time independent. The BC framework presented in Section \ref{Sec:BCs} can be applied in the case of a different gauge choice, although the characteristic structure of the evolution equations may change in a non-trivial way. To properly model a black hole echo event, one may want some way of controlling the position of the boundary $r=a$ relative to the apparent horizon. For example, one might require that the reflection surface remain a constant proper distance from the apparent horizon, so that a pulse can be reflected without the concern that the apparent horizon would move into the domain. Similar conditions are considered in \cite{PhysRevD.74.104006,PhysRevD.70.104007}, where gauge conditions that ensure the inner boundary of the domain stays just inside the apparent horizon are used, which could be adapted to our case where we want the boundary just \emph{outside} the apparent horizon. We leave a better understanding of how to control the boundary $r=a$ to future work, that may be prescribed by yet-unknown quantum gravitational processes.

This BC framework may also be applied in a full 3D simulation. We considered here in spherical symmetry the Misner-Sharp mass as a way to measure the quasi-local mass-energy of the spacetime, but this is not a unique choice. In 3 dimensions, one would need to choose a quasi-local mass-energy measure (e.g. those found in \cite{Szabados2009}) and derive BCs such that it can be conserved in the same sense as what we have done here. Future work may include a 3D generalization of this framework.

\begin{acknowledgments}

We are thankful for valuable discussions with David Brown, Luis Lehner, and the Strong Gravity group at the Perimeter Institute. Numerical implementation for this project was accomplished with the Julia programming language \cite{Julia-2017} and the packages \cite{DifferentialEquations.jl-2017}. This research was funded thanks in part to the Canada First Research Excellence Fund through the Arthur B. McDonald Canadian Astroparticle Physics Research Institute, the Natural Sciences and Engineering Research Council of Canada, and the Perimeter Institute. Research at Perimeter Institute is supported in part by the Government of Canada through the Department of Innovation, Science and Economic Development and by the Province of Ontario through the Ministry of Colleges and Universities.
\end{acknowledgments}

\appendix

\section{Enclosed Boundary Problem for Cylindrically Symmetric Gravitational Waves}\label{App:GWs}

Black hole echoes have been discussed in the literature in the context of both scalar waves and gravitational waves (GWs). Defining reflecting boundary conditions (BCs) in the context of a local mass-energy for scalar waves was done in Section~\ref{Sec:BCs}, but this can also be done for GWs. There are no GWs in the restriction of spherical symmetry, however there are well known examples in the framework of Einstein-Rosen waves \cite{EINSTEIN193743,stephani_kramer_maccallum_hoenselaers_herlt_2003}. These vacuum solutions use coordinates $x^\mu=(t,\rho,z,\varphi)$ and have the line element
\begin{align}
    ds^2 &= e^{2(\gamma-\psi)}\left(-dt^2+d\rho^2\right)\nonumber\\&+e^{2\psi}dz^2+\rho^2e^{-2\psi}d\varphi^2\,,
\end{align}
where $\psi$ and $\gamma$ are functions of $\rho$ and $t$ only. Einstein's vacuum field equations dictate that $\psi$ satisfies the cylindrically symmetric wave equation:
\begin{align}
    \partial_t^2\psi&=\partial_\rho^2\psi+\frac{1}{\rho}\partial_\rho\psi\,,
\end{align}
and that the time dependence of $\gamma$ is dictated by
\begin{align}
    \partial_t\gamma&=2\rho\partial_t\psi\partial_\rho\psi\,,\label{Eq:gammat}
\end{align}
and is subject to the constraint
\begin{align}
    \partial_\rho\gamma=\rho\left(\partial_\rho\psi\right)^2+\rho\left(\partial_t\psi\right)^2\,.\label{Eq:gammarho}
\end{align}
The local mass-energy of the spacetime is measured with the so-called $C$-energy \cite{stephani_kramer_maccallum_hoenselaers_herlt_2003}, which is simply given by
\begin{align}
    M(\rho,t)=\gamma/4\,,
\end{align}
and thus Eqns.~(\ref{Eq:gammat}) and~(\ref{Eq:gammarho}) are proportional to its $\rho$ and $t$ derivatives. We can then write the conservation law 
\begin{align}
    \partial_t\int_a^b\frac{1}{2}\left[\left(\partial_\rho\psi\right)^2+\left(\partial_t\psi\right)^2\right]\rho\,d\rho= \rho\partial_t\psi\partial_\rho\psi\Big|_a^b\,,
\end{align}
which happens to be identical to the conservation law for the cylindrically symmetric wave equation outside of the context of general relativity. The BCs that keep the $C$-energy from changing are then given by the classical Dirichlet/Neumann conditions $\partial_t\psi=0$ and $\partial_\rho\psi=0$. We can then define a reflecting IBVP as we did in spherical symmetry. The BCs are simple due to the gauge fixed assumption of the line element considered here, but this may not be the case if a more general line element where the ``perimeter radius'' $\rho$ is allowed to be dynamic, as it was in our dynamic examples in spherical symmetry.

\section{Static scalar field Summation By Parts}\label{App:SBP}

The Klein-Gordon equation $\nabla^\mu\nabla_\mu\phi-m^2\phi=0$ can be reduced to a first order system with the definition of two different auxiliary variables than those considered in the main text: $\Pi\equiv\partial_t\phi$ and ${\psi^i\equiv[(\alpha^2-\beta_j\beta^j)\partial^i\phi + \beta^i\partial_t\phi]/\alpha}$. The evolution equations in spherical symmetry for a static background look particularly simple,
\begin{align}
\partial_t \phi&= \Pi\,,\\
\partial_t\psi^r&=\beta^r\nabla_r\psi^r +\alpha\partial^r\Pi-m^2\alpha\beta^r\phi\,,\\
\partial_t \Pi&= \alpha\nabla_r\psi^r+\beta^r\partial_r\Pi-m^2\alpha\phi\,,
\end{align}
where $\nabla_r\psi^r=(\sqrt{\gamma})^{-1}\partial_r(\sqrt{\gamma}\,\psi^r)$ is the 3-divergence. This system of equations admits the following conservation law as long as $c_+c_-<0$ throughout the domain:
\begin{align}
    &\partial_t\int_a^b\frac{1}{2}\left(\frac{\alpha\Pi^2+\alpha\psi_r\psi^r-2\psi_r\beta^r\Pi}{(\alpha^2-\beta_r\beta^r)}+m^2\phi^2\right)\sqrt{\gamma}\,dr\nonumber\\&= \Pi\psi^r\sqrt{\gamma}\Big|^b_a\,.
\end{align}
Distributing the time derivative and substituting the evolution equations, we obtain 
\begin{align}
    \int^b_a\Pi\nabla_r\psi^r\sqrt{\gamma}\,dr+\int^b_a(\partial_r\Pi)\psi^r\sqrt{\gamma}\,dr=\Pi\psi^r\sqrt{\gamma}\Big|_a^b\,. \label{Eq:CovScalar}
\end{align}
Crucially, the definition of the auxiliary variables $\Pi$ and $\psi^r$ make the flux term on the right hand side a simple multiple of the two variables and Eq.~(\ref{Eq:CovScalar}) is in exactly the same form as Eq.~(\ref{Eq:CovariantIntegration}), which will set up the problem to be directly mimicked by the covariant SBP property when $\Pi$ and $\psi^r$ become grid functions:
\begin{align}
\Pi^\mathrm{T}W \nabla \psi^r + (D \Pi)^\mathrm{T}W \psi^r  = \Pi \psi^r\sqrt{\gamma}\big|_a^b\,,
\end{align}
where the operator $\nabla \equiv \Gamma^{-1}D\Gamma$ approximates the covariant 3-divergence of 3-vectors, the operator $D$ approximates the covariant scalar gradient, and $W\equiv\Sigma\Gamma$ is the covariant norm operator and approximates covariant integration.

\bibliography{refs}

\end{document}